\documentclass[prd,twocolumn,showpacs,preprintnumbers,amsmath,nofootinbib,amssymb,10pt]{revtex4}

\usepackage{graphics}
\usepackage[utf8]{inputenc}
\usepackage{epsfig}
\usepackage{subfigure}
\usepackage{dcolumn}
\usepackage{bm}
\usepackage{color}

\begin{document}

\title{Radiative decays of the neutral $Z_c(3900)$ and $Z_c(4020)$}

\author{Xiao-Yun Wang$^1$, Gang Li$^1$~\footnote{ gli@qfnu.edu.cn}, Chun-Sheng An$^2$~\footnote{ancs@swu.edu.cn}, Ju-Jun Xie$^{3,4,5}$}

\affiliation{$^1$College of Physics and Engineering, Qufu Normal University, Qufu 273165, People's Republic of China}
\affiliation{$^2$School of Physical Science and Technology, Southwest University, Chongqing 400715, China}
\affiliation{$^3$Institute of Modern Physics, Chinese Academy of Sciences, Lanzhou 730000, China} 
\affiliation{$^4$School of Nuclear Science and Technology, University of Chinese Academy of Sciences, Beijing 101408, China}
\affiliation{$^5$Lanzhou Center for Theoretical Physics, Key Laboratory of Theoretical Physics of Gansu Province, Lanzhou University, Lanzhou, Gansu 730000, China}

\begin{abstract}

We study the radiative decays $Z_c(3900)/Z_c(4020) \to \gamma \chi_{cJ}(\gamma\chi_{cJ}^\prime)$ ($J=0, 1, 2$), with the assumption that the $Z_c(3900)$ and $Z_c(4020)$ couple strongly to $D\bar D^* +c.c$ and $D^*{\bar D}^*$ channel, respectively. By considering the contributions of intermediate
charmed mesons triangle loops within an effective Lagrangian approach, it is shown that the calculated partial widths of $Z_c(3900) \to \gamma \chi_{cJ}$ are about a few hundreds keVs, while the obtained partial widths $Z_c(4020) \to \gamma \chi_{cJ}$ are about tens of keVs. The predicted partial widths of $Z_c(3900)\to\gamma\chi_{c0,1}^\prime$ are less than 1 keV, which mainly due to the very small phase space. For $Z_c(4020)\to\gamma\chi_{c0,2}^\prime$, the calculated partial widths are usually smaller than 1 keV. For the $Z_c(4020)\to\gamma\chi_{c1}^\prime$ process, the obtained partial widths can reach up to the order of 10 keV. Furthermore, the dependence of these ratios between different decay modes on the masses of $Z_c(3900)$ or $Z_c(4020)$ are also investigated, which may be a good quantity for the experiments. It is hoped that these calculations here could be tested by future experiments.

\end{abstract}

\maketitle

\section{Introduction}
\label{sec:introduction}

The discovery of $X(3872)$ in 2003 opened the gate to the abundance of the $XYZ$ structures in the
heavy quarkonium region~\cite{Belle:2003nnu}. Many of them cannot be accommodated in the conventional quark model as $Q\bar Q$ and thus turn out to be excellent candidates for exotic state. A large amount of experimental and theoretical studies are devoted to those $XYZ$ states~\cite{Chen:2016qju,Chen:2016spr,Esposito:2016noz,Guo:2017jvc,Olsen:2017bmm,Liu:2019zoy,Brambilla:2019esw,Guo:2019twa}. Among these states, the charged charmoniumlike states $Z_c(3900)$~\cite{BESIII:2013ris,Belle:2013yex} and $Z_c(4020)$~\cite{BESIII:2013ouc} have attracted special attention due to their four-quark nature. In 2013, the BESIII Collaboration first observed a new charged state $Z_c(3900)$ in the $\pi^\pm J/\psi$ invariant mass spectra of the $e^+e^-\to\pi^+\pi^-J/\psi$ reaction~\cite{BESIII:2013ris}, and it was confirmed by the Belle Collaboration in the same process~\cite{Belle:2013yex}. Later, the charged $Z_c(3900)$ is also observed in the invariant mass spectrum of $D\bar D^*$ in the open charm process $e^+e^-\to\pi^\pm(D\bar D^*)^\mp$~\cite{BESIII:2013qmu}.
In addition, the first evidence of the neutral $Z_c(3900)$ decaying into $J/\psi\pi^0$, was reported in Ref.~\cite{Xiao:2013iha} using the CLEO-c data. While the BESIII Collaboration reported the observation of neutral $Z_c(3900)$ in the $e^+e^-\to\pi^0\pi^0J/\psi$~\cite{BESIII:2015cld} and $e^+e^- \to \pi^0(D{\bar D}^*)^0$~\cite{BESIII:2015ntl}, respectively. Through the partial wave analysis of process $e^+e^-\to\pi^+\pi^-J/\psi$, the quantum numbers of $Z_c(3900)$ are determined as $I^G(J^{PC})=1^+(1^{+-})$~\cite{BESIII:2017bua,Workman:2022ynf}. 

The BESIII Collaboration found charged $Z_c(4020)$ state in the invariant mass spectrum of $\pi^\pm h_c$ in the process of $e^+e^-\to\pi^+\pi^-h_c$~\cite{BESIII:2013ouc}. The charged $Z_c(4020)$ states were confirmed in the $D^*\bar D^*$ invariant mass spectrum of the $e^+e^-\to\pi^\pm(D^*\bar D^*)^\mp$ reaction~\cite{BESIII:2013mhi}. While its neutral partner was reported in the $\pi^0h_c$ invariant mass spectrum of the $e^+e^-\to\pi^0\pi^0h_c$ reaction~\cite{BESIII:2014gnk} and in the $(D^{*} \bar{D}^{*})^{0}$ invariant mass spectrum of the $e^{+} e^{-} \to (D^{*} \bar{D}^{*})^{0} \pi^0$ reaction~\cite{BESIII:2015tix}, respectively. 

After the observation of the $Z_c(3900)$ and $Z_c(4020)$ states, many theoretical discussions have been carried out to explore their inner structures~\cite{Wang:2013cya,Aceti:2014uea,Guo:2013sya,Cui:2013yva,Zhang:2013aoa,Chen:2013omd,Soleymaninia:2013cxa,Braaten:2013boa,Faccini:2013lda,Wang:2013llv,Qiao:2013dda,Wang:2022clw,Chen:2013coa,Swanson:2014tra,Szczepaniak:2015eza,Wilbring:2013cha,Dong:2013iqa,Dong:2013kta,Li:2014pfa,Gutsche:2014zda,Esposito:2014hsa}. Since the measured masses of the $Z_c(3900)$ and $Z_c(4020)$ lie slightly above the $D{\bar D}^*$ and $D^*{\bar D}^*$  mass thresholds, it may indicate that these two states are good candidates of the hadronic
molecule~\cite{Wang:2013cya,Aceti:2014uea,Guo:2013sya,Cui:2013yva,Zhang:2013aoa,Chen:2013omd}. Besides the molecule explanation, these two states are also identified as tetraquark states~\cite{Soleymaninia:2013cxa,Braaten:2013boa,Faccini:2013lda,Wang:2013llv,Qiao:2013dda,Wang:2022clw}, or kinematical threshold effects~\cite{Chen:2013coa,Swanson:2014tra,Szczepaniak:2015eza}. Recently, the production and decay behaviors
of $Z_c(3900)$ and $Z_c(4020)$ states were extensively investigated in $D{\bar D}^*$ and $D^*{\bar D}^*$ hadronic molecule scenario in Refs.~\cite{Wilbring:2013cha,Dong:2013iqa,Dong:2013kta,Li:2014pfa,Gutsche:2014zda,Esposito:2014hsa}, where the theoretical calculations are consistent with the corresponding experimental measurements.

It is known that the intermediate meson loop (IML) transition is regarded as an important nonperturbative transition mechanism which has a long history~\cite{Lipkin:1986bi,Lipkin:1988tg,Moxhay:1988ri} and recently it is widely used to study the production and
decays of exotic states~\cite{Liu:2013vfa,Liu:2020orv,Guo:2013zbw,Li:2013yla,Chen:2013bha,Liu:2016xly,Li:2013xia,Chen:2015igx,Voloshin:2019ivc,He:2013nwa,Wu:2016ypc,Li:2012as}. The radiative transitions between neutral $Z_c(3900)/Z_c(4020)$ and the charmonia states are particular modes compared to the charged
$Z_c(3900)/Z_c(4020)$. The quark and antiquark in the different components of neutral $Z_c(3900)/Z_c(4020)$ can annihilate into a photon, and the
rest charm and anticharm quarks form a charmonium in the final state. In Ref.~\cite{Chen:2015igx}, the radiative decays of the $Z_c(3900) \to \gamma \chi_{c0,1}$ was studied in hadronic molecule picture using an effective Lagrangian approach. In Ref.~\cite{Voloshin:2019ivc}, the radiative transitions from the neutral exotic $Z_c(4020)$ resonance to $X(3872)$ was investigated.

Along this line, in this work, we estimate the radiative decays $Z_c(3900)/Z_c(4020)\to \gamma\chi_{cJ}(\gamma \chi_{cJ}^\prime)$ ($J=0, 1, 2$) with the non-relativistic effective field theory (NREFT) under assumption that the $Z_c(3900)$ and $Z_c(4020)$ couple strongly to $D\bar D^* +c.c$ and $D^*{\bar D}^*$ channel in $S$-wave, respectively. Based on this picture, in the present work, the quantum numbers of the neutral $Z_c(4020)$ state are taken to be $I^G(J^{PC})=1^+(1^{+-})$, which are consistent with those previous theoretical results, for instance, in Refs.~\cite{Wang:2013cya,Chen:2015igx,Voloshin:2019ivc,He:2013nwa,Aceti:2014uea}.

The rest of this paper is organized as follows. In Sec.~\ref{sec:framework}, we present the theoretical framework used in this work. In Sec.~\ref{sec:results}, numerical results are presented, and a brief summary is given in Sec.~\ref{sec:summary}.

\section{Theoretical Framework}    \label{sec:framework}

\subsection{Triangle diagrams}

\begin{figure*}[htbp]
\includegraphics[width=0.85\textwidth]{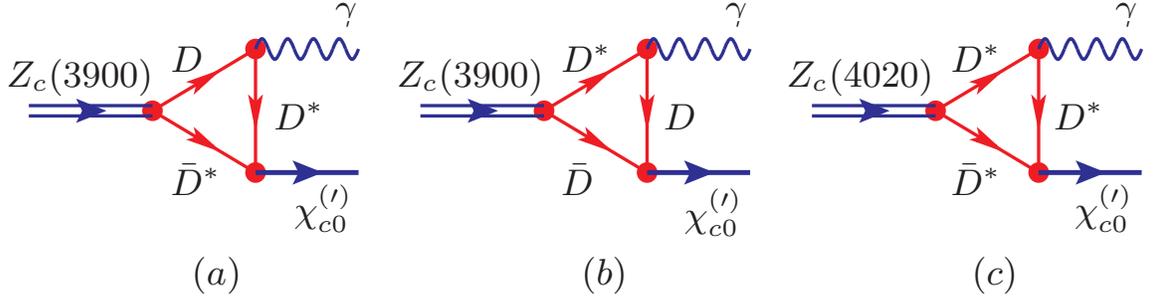}
\caption{The hadron-level diagrams of the radiative decay processes $Z_c(3900)/Z_c(4020) \to \gamma \chi_{c0}^{(\prime)}$. The charge conjugated diagrams are not shown but included in the calculations.}
\label{fig:loops}
\end{figure*}

We study the radiative decays of $Z_c(3900)$ and $Z_c(4020)$ states within the effective Lagrangian approach. Based on the strong couplings of $Z_c(3900)$ and $Z_c(4020)$ to $D\bar{D}^*$ and $D^*\bar{D}^*$, respectively, their radiative decays can be proceed via those triangle diagrams as shown in Fig.~\ref{fig:loops}, where there are three charmed mesons in the triangle loop. To be specific, we denote the one connecting the initial $Z_c(3900)/Z_c(4020)$ and the photon as $M1$, the one connecting the $Z_c(3900)/Z_c(4020)$ and the final $\chi_{cJ}^{(\prime)}$ as $M2$ and the exchanged meson between $\gamma$ and $\chi_{cJ}$ as $M3$. For example, in Fig.~\ref{fig:loops} (a), $M1$, $M2$ and $M3$ are the $D$, ${\bar D}^*$ and $D^*$, respectively. In addition, all the loops contributing to each decay  are listed in Table~\ref{tab:loops}.

\begin{table}[htbp]
\centering
\caption{ All the possible loops contributing to each transition. The mesons are listed as $[M1,M2,M3]$.}
\begin{tabular}{l l}
\hline
$Z_c(3900) \to \gamma \chi_{c0}^{(\prime)}$   & [$D$,   ${\bar D^*}$, $D^*$], [$D^*$,  ${\bar D}$,  $D$]\\
$Z_c(3900) \to \gamma \chi_{c1}^{(\prime)}$   & [$D^*$, $\bar D$, $D^*$]\\
$Z_c(3900) \to \gamma \chi_{c2}^{(\prime)}$   & [$D$,   ${\bar D}^*$, $D^*$]\\
$Z_c(4020) \to \gamma \chi_{c0}^{(\prime)}$   & [$D^*$, ${\bar D^*}$, $D^*$]\\
$Z_c(4020) \to \gamma \chi_{c1}^{(\prime)}$   & [$D^*$, $\bar D^*$, $D$]\\
$Z_c(4020) \to \gamma \chi_{c2}^{(\prime)}$   & [$D^*$, $\bar D^*$, $D^*$]\\
\hline
\end{tabular}\label{tab:loops}
\end{table}

\subsection{Effective interaction Lagrangians}
In order to calculate the triangle loops shown in Fig.~\ref{fig:loops}, we firstly need the effective couplings of the $Z_c(3900)$ and $Z_c(4020)$ states to $D\bar{D}^*$ and $D^*\bar{D}^*$ channels, respectively. The effective Lagrangians describing the couplings of $Z_c(3900)/Z_c(4020)$ to $D\bar{D}^*$ and $D^*\bar{D}^*$ channels via $S$-wave can be easily written as~\cite{Cleven:2013sq},

\begin{eqnarray}
{\cal L}_{Z_c^{(\prime)}} &=& z^\prime \epsilon^{ijk} {\bar V}_a^{\dagger i} Z_c^{\prime j} V_a^{\dagger k}  + z[ {\bar V}_a^{\dagger i} Z_c^i P_a^\dagger \nonumber\\
&&-{\bar P}_a^{\dagger } Z_c^i V_a^{\dagger i}] + {\mbox {H.C.}}
\end{eqnarray}
where $Z_c$ and $Z'_c$ stand for $Z_c(3900)$ and $Z_c(4020)$, respectively. While $V_a$ and $P_a$ are the vector and pseudoscalar charmed mesons, respectively, i.e. $P_a(V_a)=(D^{(*)0},D^{(*)+},D_s^{(*)+})$. In the two-component notation of Ref.~\cite{Hu:2005gf}, the charmed mesons are represented by $H_a = \vec{V_a}\cdot\vec{\sigma}+P_a$, with $\vec \sigma$ being the Pauli matrices, and $a$ is the light flavor index. Besides, $z$ and $z^\prime$ are the effective couplings. With the above effective Lagrangians, we can obtain

\begin{eqnarray}
\Gamma_{Z_c(3900) \to D^0\bar{D}^{*0}} &=& \frac{|z|^2}{4\pi}\frac{|\vec{q}_{D}|}{M_{Z_c(3900)}}M_D M_{D^*},\\
\Gamma_{Z_c(4020) \to D^{*0}\bar{D}^{*0}} &=& \frac{|z^\prime|^2}{4\pi}\frac{|\vec{q}_{D^*}|}{M_{Z_c^\prime(4020)}}M_{D^*}^2,
\end{eqnarray}
with $\vec{q}_D$ and $\vec{q}_{D^*}$ being the three-momenta of $D^0$ and $D^{*0}$ meson in the rest frame of $Z_c(3900)$ and $Z_c(4020)$, respectively. Here, we have assumed that the total widths of $Z_c(3900)$ and $Z_c(4020)$ are saturated by the decays $Z_c(3900) \to D^0{\bar D}^{*0}+{\bar D}^0 D^{*0}$ and $Z_c(4020) \to D^{*0}{\bar D}^{*0}$. With these values for $\Gamma_{Z_c(3900) \to  D^0{\bar D}^{*0}+{\bar D}^0 D^{*0}} = 35\pm 19$ MeV~\cite{BESIII:2015ntl} and $\Gamma_{Z_c(4020) \to D^{*0}\bar{D}^{*0}} = 23\pm 6$ MeV~\cite{BESIII:2015tix}, one can get the relevant coupling constants as follows,
\begin{eqnarray}
|z|&=& (1.67 \pm 0.45)~ {\rm GeV}^{-1/2} \, , \\
|z^\prime| &=& (0.43 \pm 0.06) ~ {\rm GeV}^{-1/2} \, .
\end{eqnarray}
where the errors are obtained with the uncertainties of the partial width of $Z_c(3900) \to  D^0{\bar D}^{*0}+{\bar D}^0 D^{*0}$ and $Z_c(4020) \to D^{*0}\bar{D}^{*0}$ decays~\cite{BESIII:2015ntl,BESIII:2015tix}. Note that to get these above coupling constants, we have used $M_{Z_c(3900)} = 3885.7$ MeV~\cite{BESIII:2015ntl}, $M_{Z_c(4020)} = 4025.5$ MeV~\cite{BESIII:2015tix}, $M_D = 1864.83$ MeV and $M_{D^*} = 2006.85$ MeV as quoted in the PDG~\cite{Zyla:2020zbs}.

On the other hand, the leading order Lagrangian for the coupling of the P-wave charmonium fields to the charmed and anticharmed mesons
can be constructed considering parity, charge conjugation conjugation, and spin symmetry~\cite{Casalbuoni:1996pg}, which can be written as
\begin{eqnarray}
{\cal L}_{\chi} &=& i\frac{g_1}{2} Tr[\chi^{\dagger i} H_a \sigma^i \bar H_a]  +H.c. \, , \label{eq:P-wave}
\end{eqnarray}
where ${\bar H}_a = -\vec{\bar V}_a \cdot \vec{\sigma} + \bar P_a$ is the anti-charmed mesons fields. $g_1$ is the coupling constant of the ground P-wave charmonia to the charmed and anticharmed mesons. The Lagrangians for the coupling of the radial excited charmonia to the charmed and anticharmed mesons have the same form as Eq.~(\ref{eq:P-wave}) with the coupling constants changed to those for the excited states $g_1^\prime$. Then, the Lagrangian of $\chi_{cJ}$ reads
\begin{eqnarray}
{\cal L}_{\chi}&=&ig_1\chi_{c2}^{\dagger ij}(V_a^i{\bar V}_a^j+V_a^j{\bar V}_a^i)+\sqrt2 g_1\chi_{c1}^{\dagger i}(V_a^i{\bar P}_a+P_a{\bar V}_a^i)\nonumber\\
&&+\frac{i}{\sqrt 3}g_1\chi_{c0}^{\dagger}(\vec{V}_a\cdot\vec{\bar V}_a+3P_a{\bar P}_a) \, ,
\label{eqs:chi}
\end{eqnarray}
where the trace and symmetry properties $\chi_{c2}^{ij}\delta^{ij}=0$ and $\chi_{c2}^{ij}\epsilon^{ijk}=0$ have been used in the derivations. In this work we take $g_1 = 4$ ${\rm GeV}^{-1/2}$ which is from an estimation using the vector meson dominance in Ref.~\cite{Colangelo:2003sa}. With $M_{\chi_{c0}^\prime}=3.869$ GeV and $\Gamma_{\chi_{c0}^\prime \to D\bar D}=22.3$ MeV calculated in LP model~\cite{Deng:2016stx,Gui:2018rvv}, we get $g_1^\prime \simeq 1.28$ ${\rm GeV}^{-1/2}$.

Finally, we also need the effective Lagrangians for these vertexes of charmed mesons and photon, which are~\cite{Amundson:1992yp,Hu:2005gf,Guo:2013zbw}
\begin{eqnarray}
{\cal L}_{\gamma} \!\!\! &=& \!\!\! \frac{e\beta}{2} Tr[H_a^{\dagger}H_b\vec{\sigma}\cdot\vec{B}Q_{ab}]  +\frac{e Q^\prime}{2m_Q}Tr[H_a^{\dagger}\vec{\sigma}\cdot\vec{B}H_a]\, , \label{eq:photon}
\end{eqnarray}
where $Q = {\rm diag} \{2/3,-1/3,-1/3\} $
is the light quark charge matrix, and $Q^\prime$ is the heavy quark
electric charge (in units of $e$).  $\beta$ is and effective coupling constant and, in this work, we take $\beta = 3.0$ GeV$^{-1}$ that is determined in the nonrelativistic constituent quark model and is adopted in the study of radiative $D^*$ decays~\cite{Amundson:1992yp}. In Eq.~(\ref{eq:photon}), the first term is the magnetic moment coupling of the light quarks, while the second one is the magnetic moment coupling of the heavy quark and hence is suppressed by $1/m_Q$.

Then, follow the works of Ref.~\cite{Cleven:2013sq}, where the decays of $Z_b(10610)/Z_b(10650) \to \gamma \chi_{bJ} (J=1,2,3)$ were investigated within the same framework as here. The transition amplitudes for $Z_c (3900)/Z_c(4020) \to \gamma \chi_{cJ}(\gamma\chi_{cJ}^\prime )$ are similar to those given in Ref.~\cite{Cleven:2013sq}, where one just needs to change bottom quark to charm quark sector. Thus the decay amplitudes of these triangle diagrams shown in Fig.~\ref{fig:loops} can be easily obtained. We present the explicit transition amplitudes for $Z_c (3900)/Z_c(4020) \to \gamma \chi_{cJ}(\gamma\chi_{cJ}^\prime )$ in the Appendix~\ref{appendix}.

Finally, the partial decay width of $Z_c (3900)/Z_c(4020) \to \gamma \chi_{cJ}(\gamma\chi_{cJ}^\prime )$ are given by

\begin{eqnarray}
\Gamma(Z_c(3900) \to \gamma \chi_{cJ}^{(\prime)}) &=& \frac {E_\gamma  |{\cal M}_{Z_c(3900) \to \gamma \chi_{cJ}^{(\prime)}}|^2 }{24\pi M_{Z_c(3900)}^2}\, , \label{eq:Gamma3900}\\
\Gamma(Z_c(4020) \to \gamma \chi_{cJ}^{(\prime)}) &=& \frac {E_\gamma^\prime |{\cal M}_{Z_c(4020) \to \gamma \chi_{cJ}^{(\prime)}}|^2}{24\pi M_{Z_c(4020)}^2}  \, , \label{eq:Gamma4020}
\end{eqnarray}
where $E_\gamma$ and $E^\prime_\gamma$ are the photon energies in the $Z_c(3900)$ and $Z_c(4020)$ rest frame, respectively.

\section{Numerical results and discussions} \label{sec:results}

The $X(3927)$ was observed in the $\gamma \gamma \to D\bar D$ process by Belle~\cite{Uehara:2005qd} and Babar~\cite{Aubert:2010ab} Collaborations, and has been a good candidate for $\chi_{c2}^\prime$ state~\cite{Li:2009zu}. In this work, we take $M_{\chi_{c2}^\prime} =3.927$ GeV as measured by the experiments~\cite{Zyla:2020zbs}. In the observed spectrum of the charmonia, the only
candidate of the $\chi_{c1}^\prime$ with well-established quantum numbers
is $X(3872)$, which was discovered by the Belle Collaboration~\cite{Belle:2003nnu}. However, since the proximity to the $D {\bar D}^*$ threshold, the interpretation of the $X(3872)$ as a molecular state or virtual state is very intriguing. The measured mass of $X(3872)$ is much lighter mass than potential quark model predictions~\cite{Deng:2016stx,Li:2009zu,Barnes:2005pb}. Thus we will not identify directly the $X(3872)$ as the $\chi_{c1}^\prime$, and the mass of the $\chi_{c1}^\prime$
will also be allowed to vary. To be specific, a range from
$3.83$ to $3.93$ GeV for the mass of $\chi'_{c1}$ and $X(3872)$ will be chosen that covers the predicted values from quark models~\cite{Deng:2016stx,Li:2009zu,Barnes:2005pb}. 

The $X(3860)$ observed in the process $e^+e^- \to J/\psi D\bar D$ by the Belle collaboration~\cite{Chilikin:2017evr} serves as a good candidate for the $\chi_{c0}^\prime$ state. The measured mass and width fit the expectation of the $\chi_{c0}^\prime$ state predicted in the
potential models~\cite{Deng:2016stx}. Thus, we will take the mass of the $\chi_{c0}^\prime$ in a range from $3800$ to $3900$ MeV which covers the predicted values of the mass of $\chi_{c0}^\prime$ from these quark models~\cite{Deng:2016stx,Li:2009zu,Barnes:2005pb} and the experimental measurements~\cite{Chilikin:2017evr}. In Table.~\ref{MchicJ}, we list the obtained charmonium masses in previous works. One can see that these values from different models are consistent with each other.

\begin{table}[htb]
\begin{center}
\caption{The final charmonium masses (in unit of GeV) are taken from the PDG~\cite{Zyla:2020zbs}, the calculated with screened potential (SP) in Ref.~\cite{Li:2009zu}, linear potential (LP) in Ref.~\cite{Barnes:2005pb}, and the results with SP and LP in Ref.~\cite{Deng:2016stx}.}\label{MchicJ}
\begin{tabular}{cccccccc}
\hline \hline
Name   & $J^{PC}$  & Exp.~\cite{Zyla:2020zbs} & \cite{Barnes:2005pb} & \cite{Li:2009zu} & LP~\cite{Deng:2016stx} & SP~\cite{Deng:2016stx}  \\ \hline
$\chi_{c0}$  & $0^{++}$  & $3.415$     & $3.424$   & $3.433$  & $3.415$ &$3.415$ \\ \hline
$\chi_{c1}$  & $1^{++}$  & $3.511$     & $3.505$   & $3.510$  & $3.516$ &$3.521$ \\ \hline
$\chi_{c2}$  & $2^{++}$  & $3.556$     & $3.556$   & $3.554$  & $3.552$ &$3.553$ \\ \hline
$\chi_{c0}^\prime$  & $0^{++}$  & $3.862?$    & $3.852$   & $3.842$  & $3.869$ &$3.848$ \\ \hline
$\chi_{c1}^\prime$  & $1^{++}$  & $...$       & $3.925$   & $3.901$  & $3.937$ &$3.914$ \\ \hline
$\chi_{c2}^\prime$  & $2^{++}$  & $3.927$     & $3.972$   & $3.937$  & $3.967$ &$3.937$ \\ \hline
\hline
\end{tabular}
\end{center}
\end{table}

In Table \ref{tab:result}, we listed the calculated partial widths of $Z_c(3900)/Z_c(4020) \to \gamma \chi_{cJ}$, which are obtained with $M_{Z_c(3900)} = 3885.7$ MeV, $M_{Z_c(4020)} = 4025.5$ MeV, $M_{\chi_{c0}} = 3414.7$ MeV, $M_{\chi_{c1}} = 3510.7$ MeV, and $M_{\chi_{c2}} = 3556.2$ MeV. It is found that these partial decay widths of $Z_c(3900) \to \gamma \chi_{cJ}$ are about a few hundreds of keVs, while the obtained partial decay widths $Z_c(4020) \to \gamma \chi_{cJ}$ are about tens of keVs.

\begin{table}[htbp]
\begin{center}
\caption{The decay widths of $Z_c(3900)/Z_c(4020)\to\gamma\chi_{cJ}$ in the unit of keV.}
\begin{tabular}{cccc}
\hline \hline
Decay mode &$\gamma\chi_{c0}$ &$\gamma\chi_{c1}$ &$\gamma\chi_{c2}$\\
\hline
$Z_c(3900)$   & 499.5 & 608.6 & 593.4\\
\hline
$Z_c(4020)$   & 33.4 & 99.4 & 108.9\\
\hline \hline
\end{tabular}\label{tab:result}
\end{center}
\end{table}

\begin{figure}[htbp]
\centering
\includegraphics[width=0.5\textwidth]{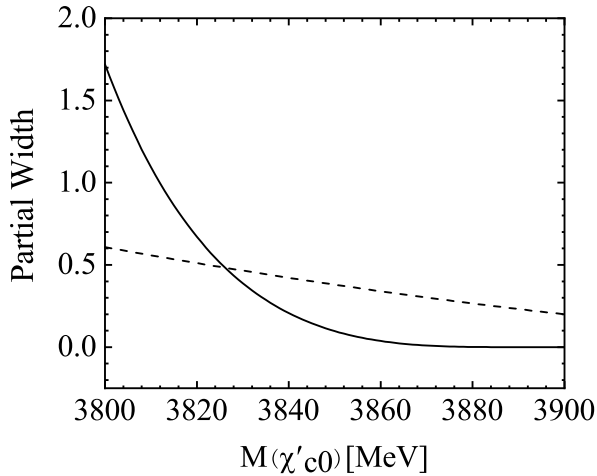}
\caption{Dependence of the decay widths of the $Z_c(3900)\to\gamma\chi_{c0}^\prime$ (solid line) and  $Z_c(4020)\to\gamma\chi_{c0}^\prime$ (dashed line) on the mass of $\chi_{c0}^\prime$.}
\label{fig:chic0prime}
\end{figure}

Next we study the mass effects of $\chi^{(\prime)}_{cJ}$ on these partial decay widths. In Fig.~\ref{fig:chic0prime}, we show the predicted partial widths of $Z_c(3900) \to \gamma \chi_{c0}^\prime$ (solid line) and $Z_c(4020) \to \gamma \chi_{c0}^\prime$ (dashed line) as a function of the mass of $\chi_{c0}^\prime$. The results show that the partial widths of $Z_c(3900)/Z_c(4020)\to\gamma\chi_{c0}^\prime$ are less than 1 keV. For the decay of $Z_c(3900) \to \gamma \chi_{c0}^\prime$ in Fig.~\ref{fig:loops}, there are two subdiagrams: $[D{\bar D}^*]D^*$ and $[D^*{\bar D}]D$. The threshold of the $D{\bar D}$ is approached at the lower end. Consequently, the curve for the width shows an increasing tendency at the lower end. While for $Z_c(4020) \to \gamma \chi_{c0}^\prime$ in Fig.~\ref{fig:loops}, only $[D^*{\bar D}^*]D^*$ subdiagram contributes. Due to  the threshold of $D^*{\bar D}^*$ is far away the range of $M_{\chi_{c0}^\prime}$ that we have chosen, the results show an monotonous behavior. Besides, with $M_{\chi_{c0}^\prime} = 3.862$ GeV obtained in Ref.~\cite{Chilikin:2017evr}, the predicted partial widths are 
\begin{eqnarray}
\Gamma(Z_c(3900) \to \gamma \chi_{c0}^\prime)&=& 0.03 \, {\rm keV}\, , \nonumber \\
\Gamma(Z_c(4020) \to \gamma \chi_{c0}^\prime)&=& 0.33 \, {\rm keV}\, .
\end{eqnarray}

\begin{figure}[htbp]
\centering
\includegraphics[width=0.5\textwidth]{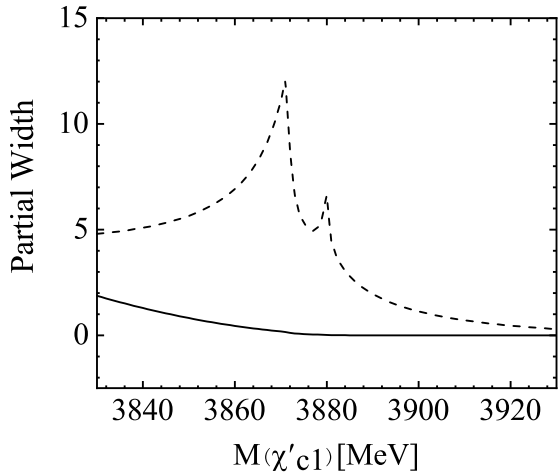}
\caption{ Dependence of the decay widths of the $Z_c(3900)\to\gamma\chi_{c1}^\prime$ (solid line) and  $Z_c(4020)\to\gamma\chi_{c1}^\prime$ (dashed line) on the mass of the $\chi_{c1}^\prime$. }
\label{fig:chic1prime}
\end{figure}

In Fig.~\ref{fig:chic1prime}, we show the numerical results for the partial widths of $Z_c(3900) \to \gamma \chi_{c1}^\prime$ (solid line) and $Z_c(4020) \to \gamma \chi_{c1}^\prime$ (dashed line) as a function of the mass of $\chi_{c1}^\prime$. The predicted partial widths of $Z_c(3900)\to\gamma\chi_{c1}^\prime$ are only a few KeVs, while the partial widths of $Z_c(4020)\to\gamma\chi_{c1}^\prime$ can reach up to $12$ KeV.
For the $Z_c(3900) \to \gamma \chi_{c1}^\prime$ process, the $[D^*{\bar D}]D^*$ subdiagram contributes, and it has very small phase space. While for the $Z_c(4020) \to \gamma \chi_{c1}^\prime$ process, the $[D^*{\bar D}^*]D$ subdiagram contributes. From Fig.~\ref{fig:chic1prime} one can see that there are two obvious cusps near the thresholds of $D^{*0}{\bar D}^{0}$ and $D^{*+}D^{-}$, respectively.  These cusps appear due to the triangle singularities
near the thresholds of neutral and charged $D{\bar D}^*$ mesons. And if we take $M_{\chi_{c1}^\prime} = 3.872$ GeV as the mass of $X(3872)$, the predicted partial widths are 
\begin{eqnarray}
\Gamma(Z_c(3900) \to \gamma \chi_{c1}^\prime)&=& 0.11 {\rm keV}\, , \nonumber \\
\Gamma(Z_c(4020) \to \gamma \chi_{c1}^\prime)&=& 8.76 {\rm keV}\, .
\end{eqnarray}
 
\begin{figure}[htbp]
\centering
\includegraphics[width=0.5\textwidth]{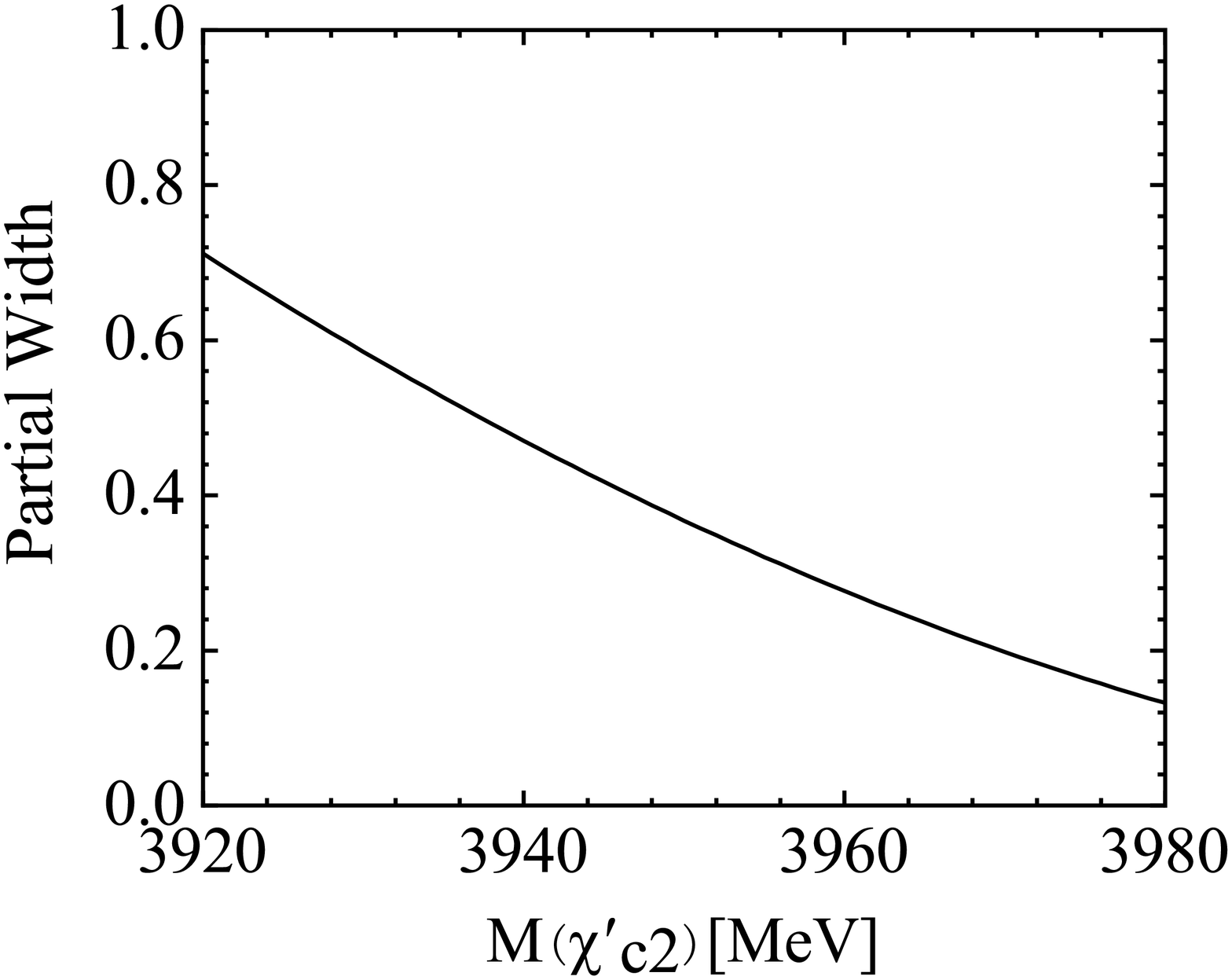}
\caption{Dependence of the decay widths of the   $Z_c(4020)\to\gamma\chi_{c2}^\prime$ on the mass of the $\chi_{c2}^\prime$.}
\label{fig:chic2prime}
\end{figure}

In Fig.~\ref{fig:chic2prime}, we present the dependence of the decay widths of the   $Z_c(4020)\to\gamma\chi_{c2}^\prime$ on the mass of the $\chi_{c2}^\prime$. The predicted partial widths of $Z_c(4020)\to\gamma\chi_{c2}^\prime$ are less than $1$ keV. The $[D^*{\bar D}^*]D^*$ subdiagram contributes this process. From Fig.~\ref{fig:chic2prime}, it is shown that the partial width monotonically decrease with the mass of $\chi_{c2}^\prime$. On the other hand, if we take $M_{\chi_{c2}^\prime} = 3.927$ GeV as the mass of the $X(3927)$ state, the predicted partial width is 
\begin{eqnarray}
\Gamma(Z_c(4020) \to \gamma \chi_{c2}^\prime)&=& 0.62 {\rm keV}\, .
\end{eqnarray}

From Eqs.~\eqref{eq:Gamma3900} and \eqref{eq:Gamma4020} we know that these partial decay widths are proportional to the product of the coupling constants $|z^{(\prime)} g_1^{(\prime)}|^2$, which will be cancelled in the ratio between different partial decay widths. Therefore, the ratios among these partial decay widths are interesting, and we define

\begin{eqnarray}
R_1 &=& \frac {\Gamma(Z_c(3900) \to \gamma\chi_{c1}^\prime)} {\Gamma(Z_c(3900) \to \gamma\chi_{c0}^\prime)} \, ,
\label{eq:ratio-Zc3900} \\
r_1 &=& \frac {\Gamma(Z_c(4020) \to \gamma\chi_{c1}^\prime)} {\Gamma(Z_c(4020) \to \gamma\chi_{c0}^\prime)} \, ,  \\
r_2 &=& \frac {\Gamma(Z_c(4020) \to \gamma\chi_{c2}^\prime)} {\Gamma(Z_c(4020) \to \gamma\chi_{c0}^\prime)} \, . \label{eq:ratio-Zc4020}
\end{eqnarray}

\begin{figure}[htbp]
\centering
\includegraphics[width=0.5\textwidth]{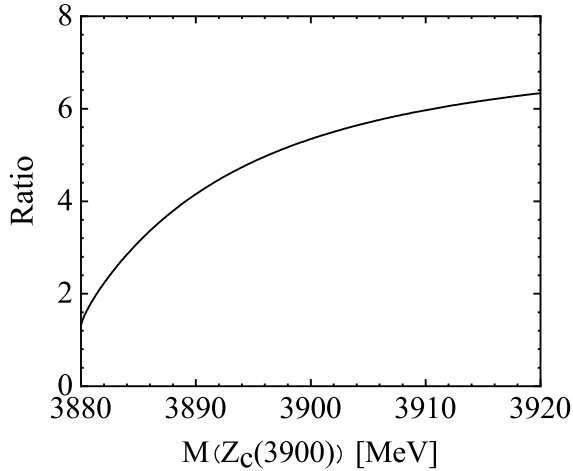}
\caption{Numerical results of $R_1$ as a function of the mass of $Z_c(3900)$.}
\label{fig:ratio_3900}
\end{figure}

The numerical results of ratio $R_1$ in terms of the mass of $Z_c(3900)$ are shown in Fig.~\ref{fig:ratio_3900}. It is seen that there is no cusp structure because the mass of $Z_c(3900)$ chosen here is above the mass threshold of $D{\bar D}^*$. In addition, the ratio is larger than one in the considered mass range of $Z_c(3900)$.

\begin{figure}[htbp]
\centering
\includegraphics[width=0.5\textwidth]{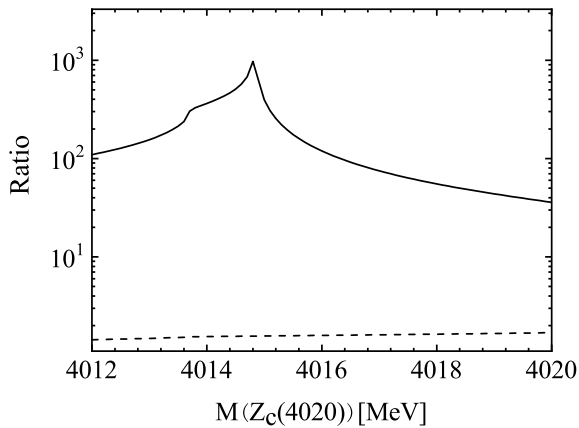}
\caption{Numerical results for $r_1$ (solid line) and $r_2$ (dashed line) as a function of the mass of $Z_c(4020)$.}
\label{fig:ratio_4020}
\end{figure}

In Fig.~\ref{fig:ratio_4020}, we present the theoretical results of the ratios $r_1$ (solid curve) and $r_2$ (dashed curve) as a function of the mass of $Z_c(4020)$. The ratio $r_1$ can reach up to about one thousand, which shows that the $Z_c(4020) \to \gamma \chi_{c1}^\prime$ is dominant and could be easier measured by experiments. For the ratio $r_1$, there is a double-cusp structure, which correspond to the thresholds of the neutral and charged $D^* {\bar D}^*$ mesons. On the other hand, the ratio $r_2$ are much less dependent on the mass of $Z_c(4020)$ and its value is about one.

\section{SUMMARY}
\label{sec:summary}

In this work, we have investigated the radiative decays $Z_c(3900)/Z_c(4020) \to \gamma \chi_{cJ}(\gamma\chi_{cJ}^\prime)$ ($J=0, 1, 2$), with the assumption that the $Z_c(3900)$ and $Z_c(4020)$ couple strongly to $D\bar D^* +c.c$ and $D^*{\bar D}^*$ channel, respectively. Including the contributions of intermediate charmed mesons triangle loops, these partial decay widths of $Z_c(3900) \to \gamma \chi_{cJ}$ and $Z_c(4020) \to \gamma \chi_{cJ}$ are calculated within an effective Lagrangian approach. With the masses quoted in the PDG~\cite{Zyla:2020zbs}, the obtained partial decay widths of $Z_c(3900) \to \gamma \chi_{cJ}$ are about of few hundreds keVs, while the obtained partial widths of $Z_c(4020) \to \gamma \chi_{cJ}$ are about tens of keVs. In addition, for the $Z_c(3900)\to\gamma\chi_{c0,1}^\prime$ and $Z_c(4020)\to\gamma\chi_{c0,2}^\prime$ decays, the partial decay width are less than 1 keV, which mainly due to the very small phase space. For the $Z_c(4020)\to\gamma\chi_{c1}^\prime$ process, the partial width can reach up to about 10 keV. Moreover, the dependence of these ratios between different decay modes on the masses of $Z_c(3900)$ or $Z_c(4020)$ are also investigated, which may be a good quantity for the experiments. These calculations here could be tested by future experiments.

\section* {Acknowledgements}

This work is supported by the National Natural Science Foundation of China, under Grants Nos. 12075133, 11835015, 11975165, 12075288, 11735003, and 11961141012. It is also partly supported by Taishan Scholar Project of Shandong Province (Grant No. tsqn202103062), the Higher Educational Youth Innovation Science and Technology Program Shandong Province (Grant No. 2020KJJ004), the Youth Innovation Promotion
Association CAS, and the
Chongqing Natural Science Foundation under Project No. cstc2021jcyj-msxmX0078.

\begin{appendix}
\section{The transition amplitudes}\label{appendix}
Here we give the amplitudes for the transitions $Z_c(3900)/Z_c(4020)\to \gamma \chi_{cJ}$. $\epsilon_1$, $\epsilon_2$ and $\epsilon_3$ are the
polarization vector of initial state, final photon and final charmonium state, respectively.

(i) $Z_c(3900) \to \gamma \chi_{c0}$
\begin{eqnarray}
\mathcal{M}_{[D,\bar{D}^*,D^*]}&=&\frac{1}{\sqrt6} e z g_1\epsilon_{ijk} \epsilon^i_1\epsilon^{j}_2 q^k \left(\beta Q + \frac{Q^\prime}{m_c}\right)\nonumber\\
&&\times I(q,D,\bar{D}^*,D^*) \, ,\nonumber\\
\mathcal{M}_{[D^*,\bar{D},D]}&=&-\frac{3}{\sqrt6} e z g_1\epsilon_{ijk} \epsilon^i_1\epsilon^{j}_2 q^k \left(\beta Q + \frac{Q^\prime}{m_c}\right)\nonumber\\
&&\times I(q,D^*,\bar{D},D) \, .
\end{eqnarray}

(ii) $Z_c(3900) \to \gamma \chi_{c1}$
\begin{eqnarray}
\mathcal{M}_{[D^*,\bar{D},D^*]}&=& -e z g_1  \left[\epsilon_1\cdot q \epsilon_2\cdot\epsilon_3-q\cdot\epsilon_3\epsilon_1\cdot\epsilon_2\right] \times \nonumber\\
&& \left(\beta Q-\frac{Q^\prime}{m_c}\right)I(q,D^*,\bar{D},D^*) \, .
\end{eqnarray}

(iii) $Z_c(3900) \to \gamma \chi_{c2}$
\begin{eqnarray}
\mathcal{M}_{[D,\bar{D}^*,D^*]}&=&- \sqrt2 e z g_1 \epsilon_{ijk} \epsilon^{il}_3 \epsilon^j_2 q^k \epsilon_{1l} \left(\beta Q + \frac{Q^\prime}{m_c}\right)\nonumber\\
&&\times I(q,D,\bar{D}^*,D^*).
\end{eqnarray}

(iv) $Z_c(4020) \to \gamma \chi_{c0}$
\begin{eqnarray}
\mathcal{M}_{[D^*,\bar{D}^*,D^*]}&=&-\frac{1}{\sqrt6} ie z^\prime g_1\epsilon_{ijk}\epsilon^i_1\left(q^k\epsilon_{2}^j-q^j\epsilon_2^k \right) \times \nonumber\\
&& \left(\beta Q-\frac{Q^\prime}{m_c}\right)
I(q,D^*,\bar{D}^*,D^*)\, .
\end{eqnarray}
(v) $Z_c(4020) \to \gamma \chi_{c1}$
\begin{eqnarray}
\mathcal{M}_{[D^*,\bar{D}^*,D]}&=& ie z^\prime g_1 \left[\epsilon_1\cdot\epsilon_2 q \cdot\epsilon_3 - \epsilon_2\cdot  \epsilon_3q\cdot\epsilon_1\right] \nonumber \\
&& \times  (\beta Q +\frac{Q^\prime}{m_c})I(q,D^*,\bar{D}^*,D).
\end{eqnarray}
(vi) $Z_c(4020) \to \gamma \chi_{c2}$
\begin{eqnarray}
\mathcal{M}_{[D^*,\bar D^*,D^*]}&=& \sqrt2 ie z^\prime g_1 \epsilon_{ijk}\epsilon^i_1\epsilon^{jl}_3 \left[q_l\epsilon^k_2- q^k\epsilon_{2l}\right] \times \nonumber \\
&&  \left(\beta Q-\frac{Q^\prime}{m_c}\right)
I(q,D^*,\bar D^*,D^*)\, .
\end{eqnarray}

In the above amplitudes, the basic three-point loop function $I(q)$ as follows~\cite{Guo:2010ak}:
\begin{widetext}
\begin{eqnarray}
I(q)&=&i\int \frac{d^d l}{(2\pi)^d}\frac{1}{(l^2-m_1^2+i\epsilon)[(P-l)^2-m_2^2+i\epsilon]} \frac{1}{[(l-q)^2-m_3^2]+i\epsilon}\nonumber\\
&=&\frac{\mu_{12}\mu_{23}}{16\pi m_1m_2m_3}\frac{1}{\sqrt a}\left[{\rm tan}^{-1}\left(\frac{c^\prime-c}{2\sqrt{ac}}\right)+{\rm tan}^{-1}\left(\frac{2a+c^\prime-c}{2\sqrt{a(c^\prime-a)}} \right)\right] \, ,
\end{eqnarray}
\end{widetext}
where the $\mu_{ij}=m_i m_j/{(m_i+m_j)}$ are the reduced masses, $b_{12}=m_1+m_2-M$, $b_{23}=m_2+m_3+q^0-M$, and the $M$ represents the mass of initial particle.
$a=\left( \mu_{23}/{m_3} \right)^2\vec{q}^ 2$, $c=2\mu_{12}b_{12}$, $c^\prime=2\mu_{23}b_{23}+\mu_{23}\vec{q}^2/{m_3}$.
$m_1$, $m_2$ and $m_3$ represent the masses of up, down and right charmed mesons in the triangle loop, respectively. 

It is worth to mention that considering the non-relativistic normalization of the charmonium and
charmed meson fields, a factor $\sqrt{M_iM_f} m_1 m_2 m_3$ should be multiplied in each amplitude. Therefore, the $I(q,M1,M2,M3)=\sqrt{M_i M_f}m_1 m_2 m_3I(q)$, where $M_i$ and $M_f$ represent the masses of initial and final particle, respectively.

\end{appendix}


\end{document}